\documentclass[12pt]{book}

\usepackage[dvips]{graphicx,color}
\usepackage{makeidx,physics,cosmology}

\makeauthorindex
\makeindex

\BookTitle{Frontier in Astroparticle Physics and Cosmology}
\CopyRight{Invited Talk at the 6th RESCEU Symposium, Nov. 4-7 2003, Tokyo, Japan.}

\begin{document}

\BookTitle{\itshape Frontier in Astroparticle Physics and Cosmology}

\pagenumbering{arabic}
\CopyRight{Invited Talk at the 6th RESCEU Symposium, Nov. 4-7 2003, Tokyo, Japan.}
\chapter{A Briefing on the Ekpyrotic/Cyclic Universe}

\author{
Justin KHOURY\\
{\it Institute for String Cosmology and Particle Astrophysics, Columbia University, New York, NY 10027, USA}\\}

\AuthorContents{J.\ Khoury}

\AuthorIndex{Khoury}{J.}

\section*{Abstract}

This is a short overview of the ekpyrotic/cyclic model of the universe, an alternative to the standard big bang inflationary paradigm.

\section{Introduction}

This truly is an exciting time for cosmology. A host of observations concord with our universe being flat and endowed with structure that grew from a nearly scale invariant, gaussian and adiabatic primordial spectrum of density perturbations. That these features coincide with the general predictions of inflation~\cite{inf} constitutes a tremendous feat for the inflationary paradigm. 

It is healthy to keep in mind, however, that since cosmology is an observational science, we will never be able to prove that cosmic acceleration did occur shortly after the big bang. Our confidence in inflation therefore cannot rest solely on observations; equally important are convincing theoretical arguments that it is the unique mechanism whose generic outcome is a universe like ours. Thus it is imperative to identify as many alternative scenarios and prove, if possible, that they are unviable or lead to different predictions that can be tested by future experiments. This endeavor is not only essential to the success of inflation; it is also good science.

The most serious candidate for a viable alternative account of cosmic history is the cyclic model of the universe~\cite{st,seiberg,tolley,design}. It proposes that time did not begin at the big bang; rather, our current period of expansion is one out of an infinite number of cycles. Each cycle consists of: (i) a hot big bang phase during which large-scale structure forms, (ii) a phase of slow, accelerated expansion which dilutes the universe, (iii) a phase of slow contraction during which nearly scale invariant density perturbations are generated, and (iv) a big crunch/bang transition at which matter and radiation are created and the next cycle is triggered. 

The model is strongly inspired by the ekpyrotic universe~\cite{ek,pert,seiberg} and shares many of its key ingredients. Most notably is the idea that the hot big bang is the result of the violent collision between two infinite branes moving along a small extra dimension. This braneworld set-up, motivated by Ho\v rava-Witten and heterotic M-theory~\cite{hw}, is crucial to ensure the safe passage of density perturbations through the bounce~\cite{pert,tolley} and to avoid chaotic mixmaster behavior~\cite{dj}.

We wish to emphasize the striking parallel between inflationary and cyclic cosmology. Here are some of the similarities that will be touched upon:
\begin{itemize}
\item Inflation and the ekpyrotic/cyclic universe can both be effectively described by a scalar field $\phi$ rolling down a potential $V(\phi)$ (Sec.~\ref{4d}). 
\item Both models are dynamical attractors and satisfy a no-hair theorem (Sec.~\ref{nohair}). 
\item The simplest inflationary and ekpyrotic/cyclic models lead to nearly identical predictions for the spectral tilt of the spectrum of density perturbations~\cite{tilt}. Moreover, these are the only two possibilities to obtain scale invariance~\cite{gratton} (Sec.~\ref{perts}).
\item The predicted spectral slopes in the two scenarios are related by an intriguing duality transformation which hints at a deeper connection (Sec.~\ref{perts}). 
\item The inflationary potential must be positive and flat; its ekpyrotic counterpart must be negative and steep. The constraints on the steepness of the latter can be described in terms of ``fast-roll'' parameters, analogous to the familiar ``slow-roll'' parameters of inflation (Sec.~\ref{frsr}).
\end{itemize}
Despite these similarities, the difference in dynamics in the two scenarios results in a key observational distinction: inflation predicts a nearly scale invariant spectrum of gravitational waves whereas the ekpyrotic model does not. 

Our analysis relies on the assumption of a successful transition across the singularity, as proposed in~\cite{seiberg}. While there has been ample literature on the subject recently~\cite{everyone}, a formal proof within string theory is still lacking. It is worth emphasizing, however, that the singularity featured in cyclic models is one of the mildest imaginable. As such it constitutes our most hopeful candidate of a singularity that could be resolved in string theory. For our purposes, we shall assume the proposal of~\cite{seiberg} for going through the bounce as well as that of Tolley {\it et al.}~\cite{tolley} for the passage of density perturbations.

\section{Four-Dimensional Effective Dynamics} \label{4d}

For most of cosmic history, except for very near the bounce, the dynamics of the cyclic model are well approximated by a four-dimensional effective action:
\begin{eqnarray}
\nonumber
& & S = \frac{M_{Pl}^2}{2}\int d^4x\sqrt{-g}\left(\frac{{\cal R}}{2}-
\frac{(\partial\phi)^2}{2}-V(\phi) + \beta^4(\phi)(\rho_M + \rho_R) \right)\,,
\end{eqnarray}
where $g$ is the determinant of the metric $g_{\mu \nu}$, ${\cal R}$ the corresponding Ricci scalar, and  $M_{Pl} = (8\pi G)^{-1/2}\approx 10^{18}$ GeV is the reduced Planck mass. The scalar field $\phi$ is interpreted in the higher-dimensional theory as measuring the distance $d$ between the two end-of-the-world branes. The precise relation is: $d = L\ln[\coth(-\phi/M_{Pl}\sqrt{6})]$, where $L$ is the bulk curvature scale. In particular, the brane collision ($d\rightarrow 0$), which coincides with the big bang/crunch transition, corresponds to $\phi\rightarrow -\infty$. The coupling function $\beta(\phi)$ is model-dependent, but $\beta \rightarrow {\rm exp}(-\phi/M_{Pl}\sqrt{6})$ as $\phi \rightarrow -\infty$, which ensures that the matter ($\rho_M$) and radiation ($\rho_R$) energy densities remains finite at the bounce~\cite{seiberg}. Moreover, $\beta(\phi)$ must be such that, for today's value of $\phi$, it satisfies tests of the Equivalence Principle. This can be achieved by making $d\beta/d\ln\phi$ small enough~\cite{st}, or via the chameleon mechanism~\cite{cham}.

\begin{figure}[t]
  \begin{center}
    \includegraphics[height=13pc]{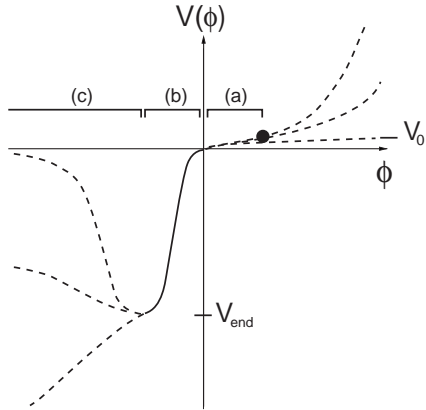}
  \end{center}
  \caption{Various allowed forms of cyclic potentials.}
\label{pot}
\end{figure}

The scalar potential $V(\phi)$, which describes the attractive force between the branes, can be divided into Regions a), b) and c). See Fig.~\ref{pot} Currently, the field lies in Region a) (indicated in the Figure with a dark circle) where the potential is flat and drives cosmic acceleration; thus $V\approx V_0\approx 10^{-120}M_{Pl}^4$. This accelerating phase makes the universe homogeneous, isotropic, flat and nearly vacuous. Eventually, the field rolls towards negative values of $V$ (Region b)), where cosmic expansion comes to a halt and the universe enters a phase of slow contraction. It is in this region that the spectrum of density perturbations is generated from quantum fluctuations in $\phi$. The field then zooms through Region c), characterized by the energy density in $\phi$ being dominated by its kinetic energy, as required by the proposal of~\cite{seiberg}. At the bounce, part of this kinetic energy is converted into matter and radiation, while the perturbations in $\phi$ are imprinted as density fluctuations in the matter/radiation fluid. Meanwhile the field rushes back to Region a) where it comes to a stop; at this time the universe enters the radiation-dominated era, marking the beginning of the next cycle.

The tightest constraints on the potential pertain to Region b), since this is where super-horizon density perturbations are generated from quantum fluctuations in $\phi$. Since $V$ is negative in this region, it follows that the equation of state $w$ for $\phi$ satisfies
\begin{equation}
w\equiv \frac{\phi'^2-2a^2V}{\phi'^2+2a^2V}> 1\,,
\label{wg1}
\end{equation}
where $a$ is the scale factor and primes denote derivatives with respect to conformal time $\tau$. It turns out that cosmology with $w>1$  has many nice properties. As shown in Sec.~\ref{nohair}, this phase of contraction is a dynamical attractor. Moreover, we will see in Sec.~\ref{perts} that $w\gg 1$ results in a spectrum of density fluctuations that is nearly scale invariant. Finally, this condition is essential to avoid chaotic mixmaster behavior at the bounce~\cite{dj}.

\section{Cosmological No-Hair Theorem} \label{nohair}

In this Section we show that accelerated expansion (as in inflation) or slow contraction (as in the cyclic model) are cosmological attractors. In other words, the dynamics in each case are highly insensitive to initial conditions and converge to the attractor after a few e-folds of contraction or expansion. 

Starting with the expanding case, consider the general Friedmann law:
\begin{equation}
3H^2M_{Pl}^2 = \frac{C_M}{a^3} + \frac{C_R}{a^4} -\frac{k}{a^2} + \frac{C_\sigma}{a^6} + \ldots + \Lambda\,,
\label{fried1}
\end{equation}
where the terms on the right-hand side describe matter ($\sim a^{-3}$), radiation ($\sim a^{-4}$), curvature ($\sim a^{-2}$), anisotropy or coherent energy of some field ($\sim a^{-6}$), and vacuum energy ($\sim a^{0}$), respectively. As the scale factor $a$ increases in time, the terms having non-zero powers of $a$ in the denominator redshift away and quickly become subdominant to the $\Lambda$ term, which remains constant. Hence, after a few e-folds, the Friedmann law becomes $3H^2M_{Pl}^2\approx \Lambda$, and the universe inflates. 

Now let us consider the contracting case. If the energy content were once again described by Eq.~(\ref{fried1}), the term that would eventually dominate would be the one with the highest power of $a$ in the denominator, since $a$ decreases in time in this case. According to Eq.~(\ref{fried1}), this would be the anisotropy term proportional to $a^{-6}$. This would indicate that the approximation of homogeneity and isotropy breaks down sometime before the bounce.

In the ekpyrotic/cyclic scenario, however, one must also include the energy density in $\phi$, $\rho_{\phi}\sim a^{-3(1+w)}$:
\begin{eqnarray}
\nonumber
3H^2M_{Pl}^2 &=& \frac{C_M}{a^3} + \frac{C_R}{a^4} -\frac{k}{a^2} + \frac{C_\sigma}{a^6} + \ldots + \Lambda\\
&+& \frac{C_{\phi}}{a^{3(1+w)}}\,.
\label{fried2}
\end{eqnarray}
From Eq.~(\ref{wg1}), $w$ is greater than unity, and so $3(1+w)> 6$. Hence, in this case it is not the anisotropy term that dominates as $a\rightarrow 0$, but instead the energy density in $\phi$. In other words, the ekpyrotic/cyclic period of slow contraction is a cosmological attractor. Moreover, this implies that the evolution becomes very simple as $a\rightarrow 0$; here the approximation of spatial flatness, homogeneity and isotropy only becomes better in time. The dynamics become ultra-local~\cite{dj}, thus providing hope for a safe passage through the big crunch singularity.

\section{Density Perturbation Spectrum} \label{perts}

We now turn to the generation of density perturbations in a general background cosmology. The only assumption is that the equation of state $w$ is nearly constant as the relevant range of modes exits the horizon. It is convenient to introduce $\bar{\epsilon}\equiv (3/2)(1+w)$, which is also nearly constant. Keep in mind that $\bar{\epsilon}$ reduces to the usual slow-roll parameter, $\epsilon_s\equiv M_{Pl}^2V_{,\phi}^2/2V^2$, in the slow-roll limit. In terms of $\bar{\epsilon}$, the scale factor for the background cosmology is given by
\begin{equation}
a(\tau) \sim (-\tau)^{1/(\bar{\epsilon}-1)}\,,
\label{a}
\end{equation}
where conformal time $\tau$ is negative and increases towards zero.

Metric perturbations are conveniently described in terms of the gauge invariant variable $u$, related to the Newtonian potential $\Phi$ by $u\equiv a\Phi/\phi'$. The Fourier mode $u_k$ with wavenumber $k$ then satisfies~\cite{gratton}
\begin{equation}
u_k'' + k^2u_k - \frac{1}{\tau^2(1-\bar{\epsilon})^2}\left\{\bar{\epsilon} -
\frac{(1-\bar{\epsilon}^2)}{2}\frac{d\ln\bar{\epsilon}}{d{\cal N}} 
\right\} u_k = 0\,, 
\label{u2}
\end{equation}
where the dimensionless time variable ${\cal N}$ is defined by $d{\cal N} \equiv -d\ln(aH)$.

Eq.~(\ref{u2}) can be solved analytically~\cite{gratton}, and the deviation from scale invariance is readily obtained:
\begin{equation}
n_s - 1 \approx  -\frac{2}{(1-\bar{\epsilon})^2}\left\{\bar{\epsilon} -
\frac{(1-\bar{\epsilon}^2)}{2}\frac{d\ln\bar{\epsilon}}{d{\cal N}} \right\}\,.
\label{ns}
\end{equation}
Using this expression, we can then identify the regimes under which the spectrum is nearly scale invariant, {\it i.e.}, $|n_s-1|\ll 1$. This is clearly satisfied when $\bar{\epsilon} \ll 1$, that is, when $w \approx -1$, corresponding to the case of slow-roll inflation. Note, however, that there is a second regime, namely $\bar{\epsilon} \gg 1$, corresponding to $w \gg 1$. This is the limit relevant to Region b) for the ekpyrotic/cyclic scenario. From Eq.~(\ref{a}), we see that the universe is expanding in the first case and contracting in the second. In Gratton {\it et al.}~\cite{gratton}, it was argued rigorously that these are the only two possibilities to obtain scale invariance.

From Eq.~(\ref{ns}), the spectral index in the inflationary limit, $\bar{\epsilon}\ll 1$, is~\cite{wang,gratton}
\begin{equation}
(n_s - 1)_{inf} \approx -2\bar{\epsilon} + \frac{d\ln\bar{\epsilon}}{d{\cal N}}\,,
\label{nsinf}
\end{equation}
while the corresponding expression for the ekpyrotic limit, $\bar{\epsilon} \gg 1$, is given by
\begin{equation}
(n_s -1)_{ek} \approx -\frac{2}{\bar{\epsilon}} - \frac{d\ln\bar{\epsilon}}{d{\cal N}}\,.
\label{nsek}
\end{equation}
The similarity of these two expressions is striking. In fact, it is easily seen that they can be transformed into one another under the map $\bar{\epsilon} \rightarrow 1/\bar{\epsilon}$. Furthermore, noting from Eq.~(\ref{a}) that $a \propto t^{1/\bar{\epsilon}} \propto H^{-1/\bar{\epsilon}}$, where $t$ is proper time, we see that inflation ($\bar{\epsilon}\ll 1$) has $a$ rapidly varying and $H$ nearly constant, whereas the ekpyrotic/cyclic model ($\bar{\epsilon}\gg 1$) has $H$ varying and $a$ nearly constant. This suggests an interesting duality between the inflationary and ekpyrotic/cyclic models that reflects itself in the final results. The invariance under $\bar{\epsilon}  \rightarrow 1/\bar{\epsilon}$ actually holds for arbitrary $w$, and for both growing and decaying modes~\cite{latham}.

\section{Fast-Roll versus Slow-Roll} \label{frsr}

We have seen in the previous section that a scale invariant spectrum of density perturbations will obtain if the universe is slowly contracting with $w$ being much larger than unity and almost constant. Equivalently, this requires $\bar{\epsilon}\ll 1$ and $d\ln\bar{\epsilon}/d{\cal N}\ll 1$. As shown in Gratton {\it et al.}~\cite{gratton}, these translate into the following conditions on Region b) of the cyclic potential $V(\phi)$:
\begin{eqnarray}
\nonumber
V &<& 0\\
\nonumber
\epsilon &\equiv& \frac{1}{M_{Pl}^2}\left(\frac{V}{V_{,\phi}}\right)^2 \ll 1 \\
|\eta| &\equiv& \left\vert 1-\frac{VV_{,\phi\phi}}{V_{\phi}^2}\right\vert\ll 1\,.
\end{eqnarray}
Qualitatively, these require, respectively, that $V$ be negative, very steep and nearly exponential in form. 
They are almost the exact opposite of the corresponding conditions for slow-roll inflation, which require the potential to be positive and nearly flat. In particular, $\epsilon$ and $\eta$ should be thought of as ``fast-roll'' parameters, in analogy with the inflationary slow-roll parameters: $\epsilon_s\equiv M_{Pl}^2V_{,\phi}^2/2V^2$ and $\eta_s\equiv M_{Pl}^2V_{,\phi\phi}/V$.

The spectral index in Eq.~(\ref{nsek}) can be neatly rewritten in terms of the fast-roll parameters~\cite{gratton}:
\begin{equation}
(n_s -1)_{ek} \approx -4(\epsilon+\eta)\,.
\end{equation}
Comparing this with the familiar expression for slow-roll inflation~\cite{lyth}, $(n_s - 1)_{inf} = -6\epsilon_s + 2\eta_s$, we see that in both inflationary and ekpyrotic/cyclic cosmology, the spectral index can be expressed simply in terms of the fast-roll or slow-roll parameters. Current observations constrain it to lie in the range $|n_s-1| < 0.1$. The above analysis shows that this requires the same degree of tuning on the potential $V(\phi)$. See~\cite{design} for a more complete analysis.

\section{Conclusion} \label{conclu}

In this note we presented a summary of the ekpyrotic/cyclic model of the universe, a recently proposed alternative account of cosmic history. We emphasized the similarities and parallels with inflationary cosmology. 
Inflation proposes that the universe underwent a short period of rapid expansion a fraction of a second after the big bang; the ekpyrotic/cyclic model proposes that the universe underwent a long period of slow contraction well before the big bang. 

Much work remains to be done to establish the ekpyrotic scenario as a serious rival to inflation. This is fair enough: inflation has been with us for over twenty years and has been the subject of thorough study, while the ekpyrotic model is still in its infancy. Inflation is based on the well-understood principles of quantum field theory, whereas many essential elements of ekpyrosis are motivated by string theory, a theory still under construction. 

This does not mean that inflation is free of unsolved problems, however. For instance, it must still address how the universe emerged from the big bang and how it entered the inflationary phase. 

The key question for ekpyrosis is whether passing through the cosmic singularity is allowed. If a bounce akin to the proposal of~\cite{seiberg,tolley,pert} is possible, then we will be faced with an important question: were the density perturbations that seeded structure formation laid out long before or shortly after the big bang? The answer will await the detection of the primordial spectrum of gravitational waves. Even if a bounce turns out not to be possible, the ekpyrotic endeavor will nevertheless have greatly increased our faith in the inflationary paradigm. 

Either outcome will constitute important progress for cosmology.

\end{document}